%% file: NuPhys2018-Ciuffoli.tex
\documentclass[12pt]{article}
\usepackage{graphicx}

\newcommand\pubnumber{NuPhys2018-Ciuffoli}
\newcommand\pubdate{\today}

\def\napoli{Institute of Modern Physics\\Chinese Academy of Sciences, Lanzhou, 730000, China}
\def\support{\footnote{EC is supported by NSFC Grants No. 11605247 and 11375201, and by the Chinese Academy of Sciences Presidents International Fellowship Initiative Grant No. 2015PM063}}

\def\Title#1{\begin{center} {\Large #1 } \end{center}}
\def\Author#1{\begin{center}{ \sc #1} \end{center}}
\def\Address#1{\begin{center}{ \it #1} \end{center}}

\newcommand\pubblock{\rightline{\begin{tabular}{l} \pubnumber\\
         \pubdate  \end{tabular}}}
\newenvironment{Abstract}{\begin{quotation}  }{\end{quotation}}
\newenvironment{Presented}{\begin{quotation} \begin{center} 
             PRESENTED AT\end{center}\bigskip 
      \begin{center}\begin{large}}{\end{large}\end{center} \end{quotation}}

\input econfmacros.tex
\begin{document}
\begin{titlepage}
\pubblock

\vfill
\Title{Measuring the Neutron Distribution from Coherent Elastic Neutrino Nucleus Scattering}
\vfill
\Author{ Emilio Ciuffoli\support}
\Address{\napoli}
\vfill
\begin{Abstract}
Last year the COHERENT collaboration was able to measure for the first time the Coherent Elastic Neutrino Nucleus Scattering (CE$\nu$NS). Neutrinos within the right energy range can be produced in large quantities at accelerator facilities via pion Decay At Rest ($\pi$DAR) and used to measure CE$\nu$NS. This new channel opens several, interesting possibilities: studying the CE$\nu$NS spectrum it will be possible, for example, to search for Physics Beyond the Standard Model, looking for deviations from the predictions of the electroweak theory; it can also give important inputs for the understanding of core collapse supernovas, where neutrino-nucleus interactions and, more generally, collective neutrino behavior play a crucial role. Using CE$\nu$NS it is also possible to measure precisely the electroweak form factor for a large number of different nuclei, extracting information on the neutron distribution inside the nucleus as well. In this presentation I will focus on the last aspect: I will calculated the precision that can be achieved in such kind of experiment, investigating in particular the effects of the low-energy threshold and the systematic errors on the quenching factor. The expected precision will be calculated using the Helm model and also with a model-independent approach.
\end{Abstract}
\vfill
\begin{Presented}
NuPhys2018, Prospects in Neutrino Physics

Cavendish Conference Centre, London, UK, December 19--21, 2018
\end{Presented}
\vfill
\end{titlepage}
\def\thefootnote{\fnsymbol{footnote}}
\setcounter{footnote}{0}

\section{Introduction}
In 2017 the COHERENT collaboration announced the first measurement of Coherent Elastic Neutrino-Nucleus Scattering (CE$\nu$NS) \cite{Coherent}. The main challenge in this kind of experiment is that the recoil energy of the nucleus (the only observable in the process) is very small, usually less than 100 keV; however thanks to recent developments in the detector technology now it is possible to measure even such a small signal. This new channel opens several interesting possibilities in the study of neutrino physics. First of all, it offers new ways to test the Standard Model: any deviation from the predictions could be a sign of new physics. Second, since neutrino interactions with nuclei and, in general, collective neutrino behavior play a crucial role in core collapse supernovas, data from CE$\nu$NS could be an input for a better understanding of this kind of phenomena; moreover, the dark matter detectors are rapidly approaching to the intrinsic neutrino floor background, and a detailed knowledge of the CE$\nu$NS cross section will be required for the next generation of detectors. Finally, from CE$\nu$NS it is possible to extract information on the electroweak form factor and on the neutron distribution inside the nucleus; in this presentation (based on the results published in \cite{mio}) we will focus on this last aspect. Neutrinos can be produced at spallation facilities via Pion Decay At Rest ($\pi$DAR): the collisions produce $\pi^\pm$ pairs; the $\pi^-$ is absorbed inside the nucleus, while the $\pi^+$ is stopped and decays at rest. At the end of the chain decay three neutrinos will be produced:
\begin{eqnarray}
\pi^+\rightarrow\mu^++\nu_{\mu}\nonumber\\
\mu^+\rightarrow e^++\nu_e+\bar{\nu}_\mu
\end{eqnarray}
Since CE$\nu$NS is a neutral current interaction, all these neutrinos will contribute in the same way to the signal and particle identification will not be possible. In Sec. \ref{Prec} we will discuss the precision that can be achieved in the measurement of the neutron distribution from CE$\nu$NS experiments, studying in particular the effect of systematic errors, such as the uncertainty on the energy reconstruction, and the low-energy threshold of the detector. The neutrino beam assumed in these calculations is the one that is being produced at China Spallation Neutron Source (CSNS), where a 1.6 GeV proton beam hits a fixed target with an average current  of 62.5 $\mu$A. Since the beam is pulsed, the steady state background can be strongly suppressed by taking into account the time structure of the beam. In order to calculate the precision on the neutron radius, the Helm model is assumed; however a model-independent way to obtain information on the neutron distribution is considered as well; these results are discussed in Sec. \ref{MIapp}.
\section{Precision on Nucleus Distribution}
\label{Prec}
\begin{figure}[h]\centering
               \includegraphics[width=0.45\textwidth]{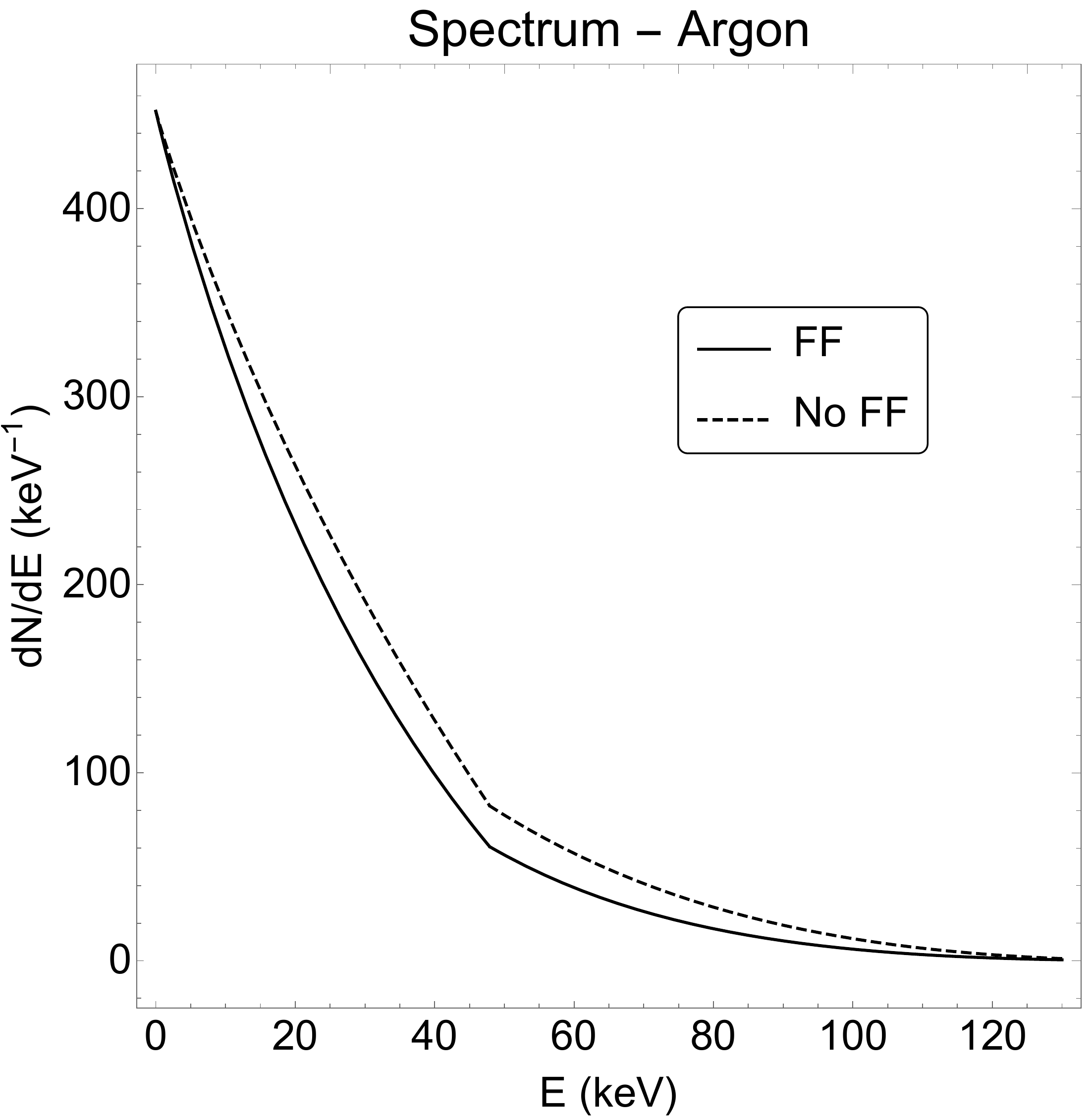}\vspace{0.2cm}
               \includegraphics[width=0.45\textwidth]{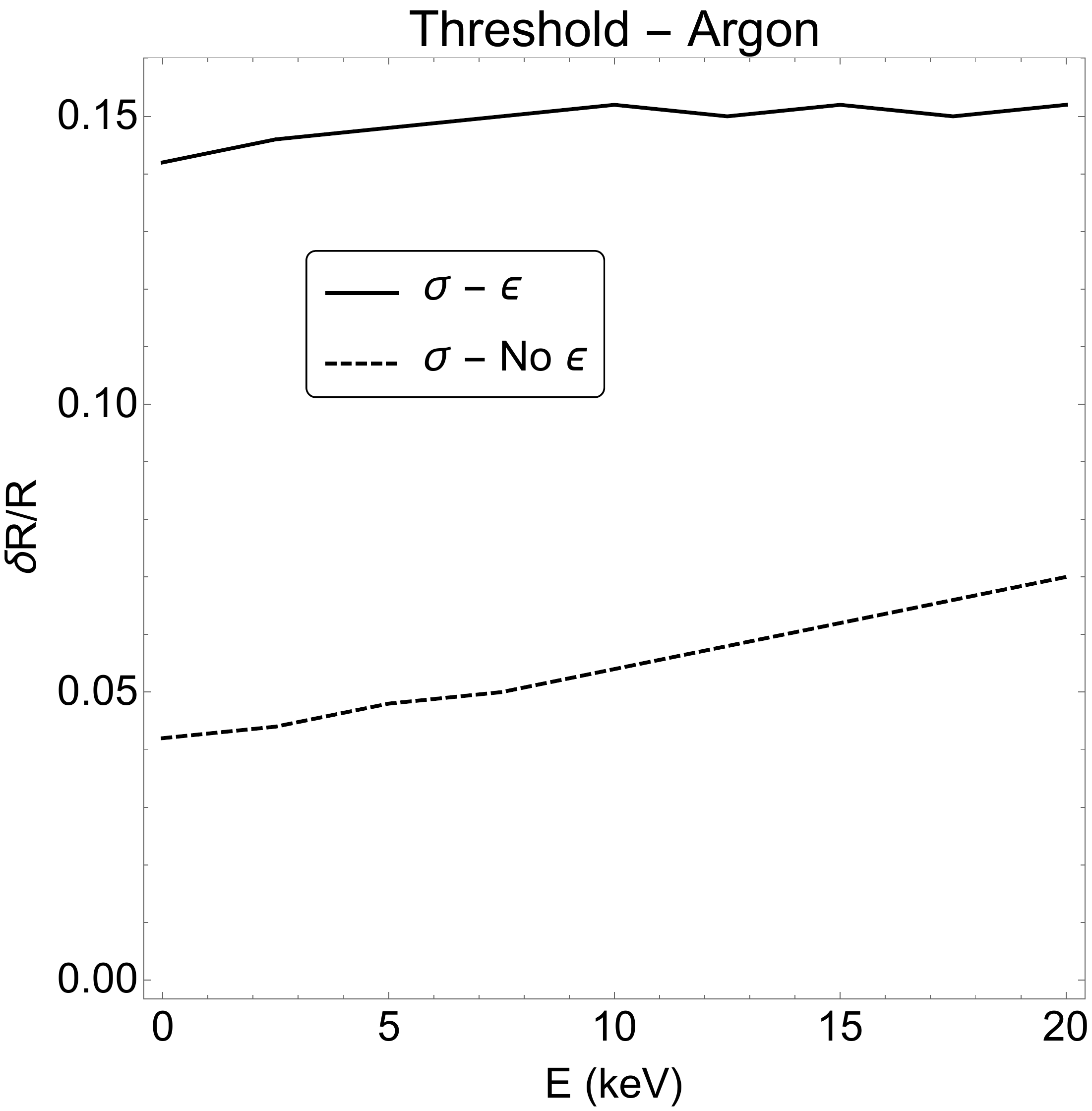}\vspace{-0.2cm}
\caption{\label{spectrum}Expected spectrum with and without the form factor (left panel), sensitivity as a function of the low-energy threshold (right panel)}
\end{figure}

From the study of CE$\nu$NS cross section it is possible to determine the electroweak form factor. Neutrinos from $\pi$DAR have energies between 0 and 55 MeV: in this energy range, the proton contribution to the form factor is strongly suppressed; for this reason from the CE$\nu$NS data it is also possible to extract information on the neutron distribution inside the nucleus (using the preliminary data from the COHERENT collaboration Cadeddu {\it et al.} obtained a first estimation of the neutron distribution radius for Cs and I \cite{Cadeddu}). Using the CSNS neutrino beam, we calculated the expected sensitivity to the neutron radius as a function of the detector mass. In order to summarize all the information on the neutron distribution in a single parameter a theoretical model must be used: in the following calculations, we used the Helm model; in principle this is a two-parameters model, since it depends on the skin thickness $s$ and the neutron bulk radius $R_n$, however, as was pointed out also in \cite{Cadeddu}, at these energies the dependence on $s$ is negligible, hence it will be considered fixed and equal to 1 fm. We considered a liquid Argon and a liquid Xenon detector,  placed at 10 m from the source and assumed 1 year lifetime. 
The most important source of systematic error is the uncertainty on the quenching factor (QF); in order to parametrize this effect we considered a simple linear model, where the relation between the observed and real energy ($E_{obs}$ and $E_{real}$, respectively) is given by
\[
E_{obs}=E_{real}(1+\epsilon)
\]
where $\epsilon$ does not depend on the energy and, when considered, it is treated as a pull parameter (see Ref. \cite{mio} for more details). 
\begin{figure}[h]\centering
               \includegraphics[width=0.45\textwidth]{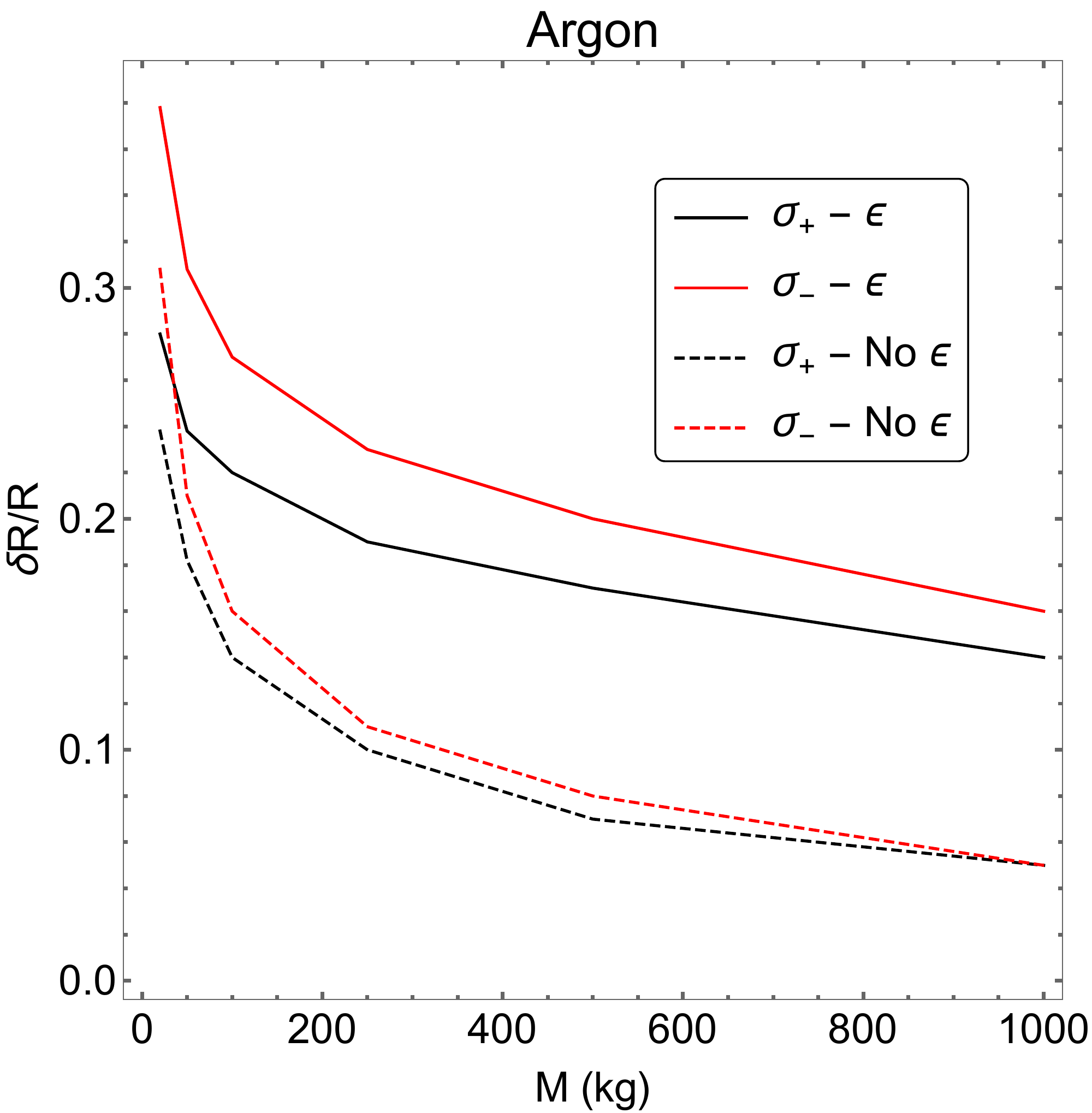}\vspace{0.2cm}
               \includegraphics[width=0.45\textwidth]{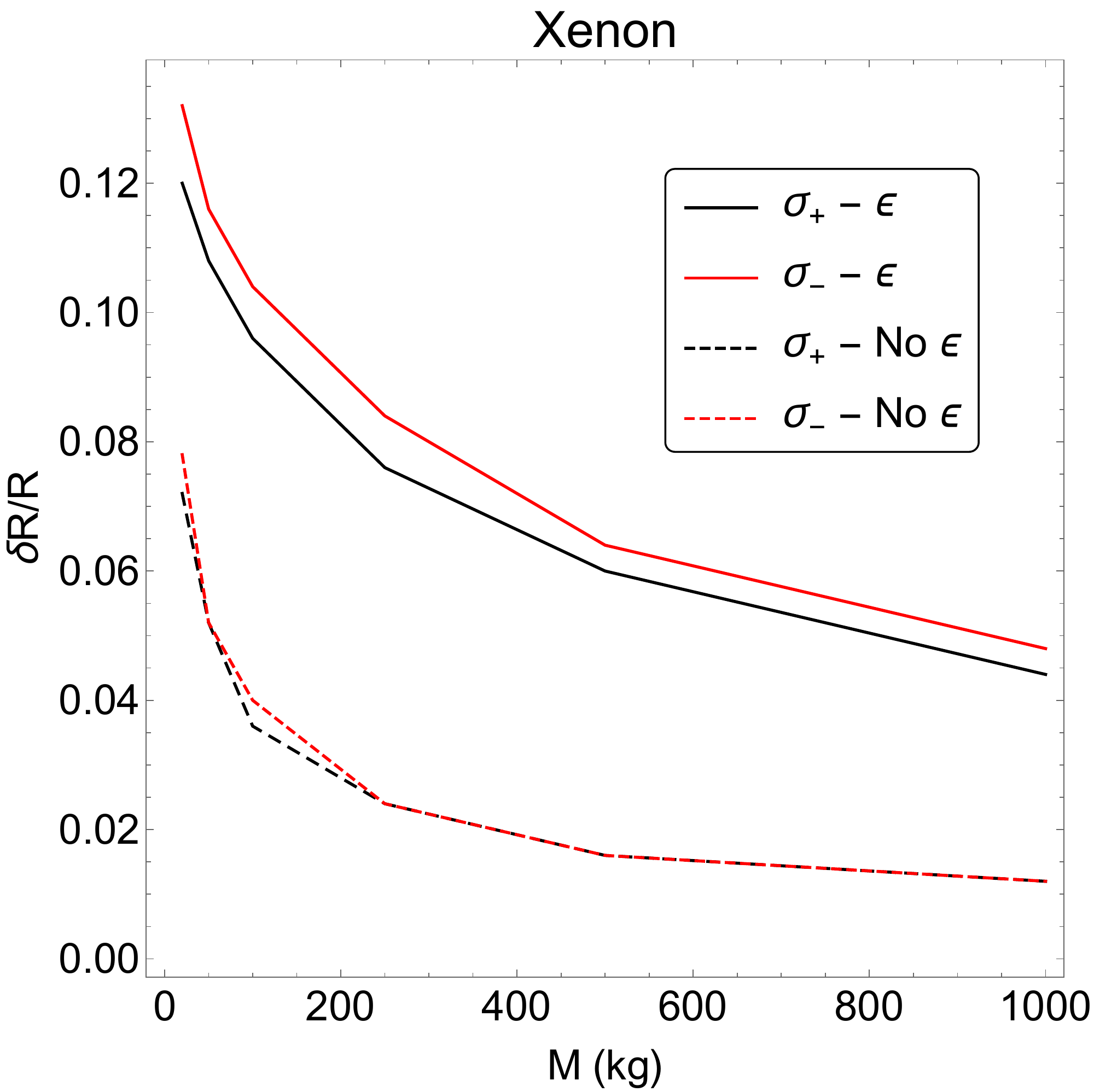}\vspace{-0.2cm}
\caption{\label{sensitivity}Expected sensitivity as a function of the detector mass}
\end{figure}
\begin{figure}[h]\centering
               \includegraphics[width=0.45\textwidth]{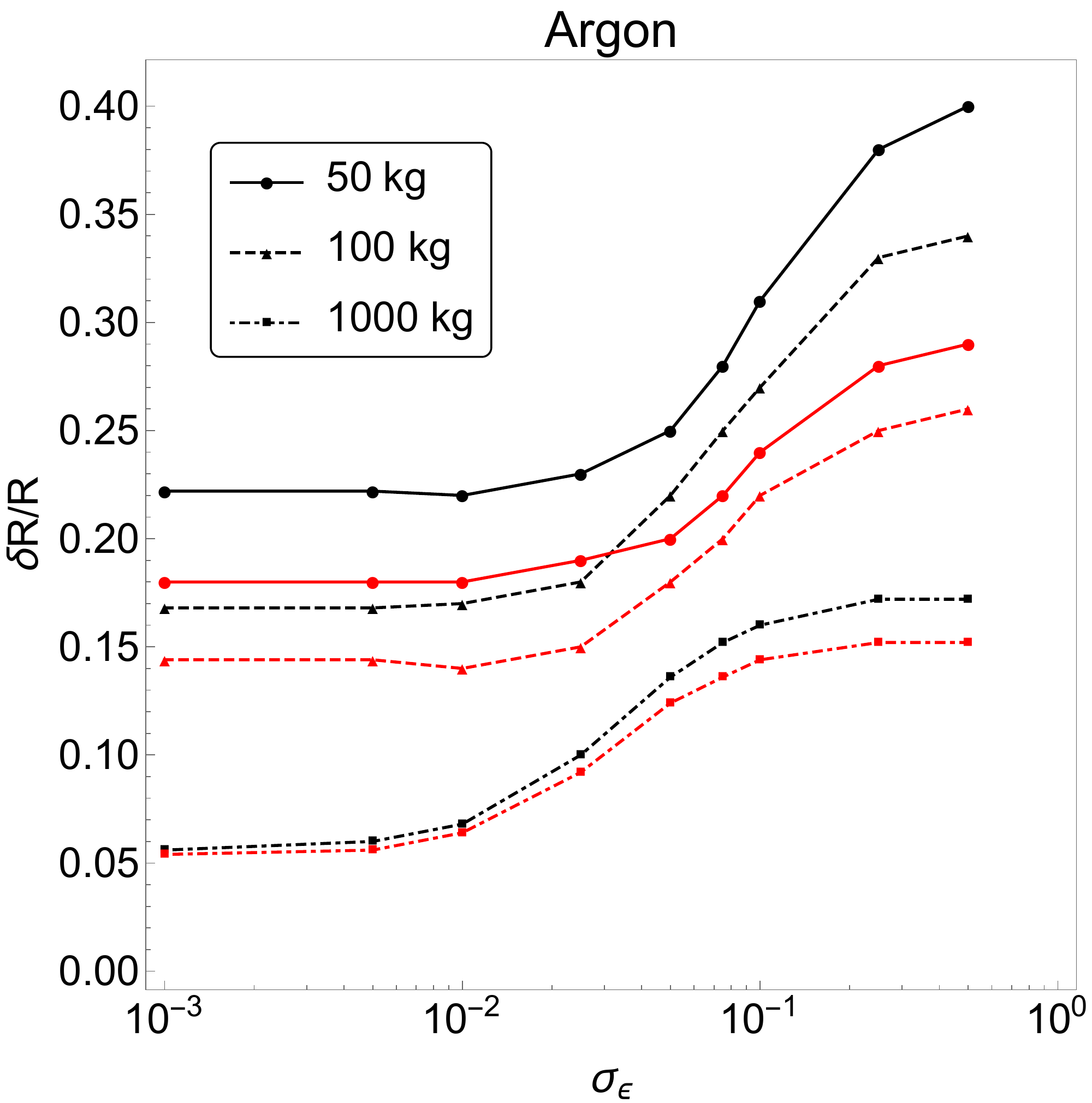}\vspace{0.2cm}
               \includegraphics[width=0.45\textwidth]{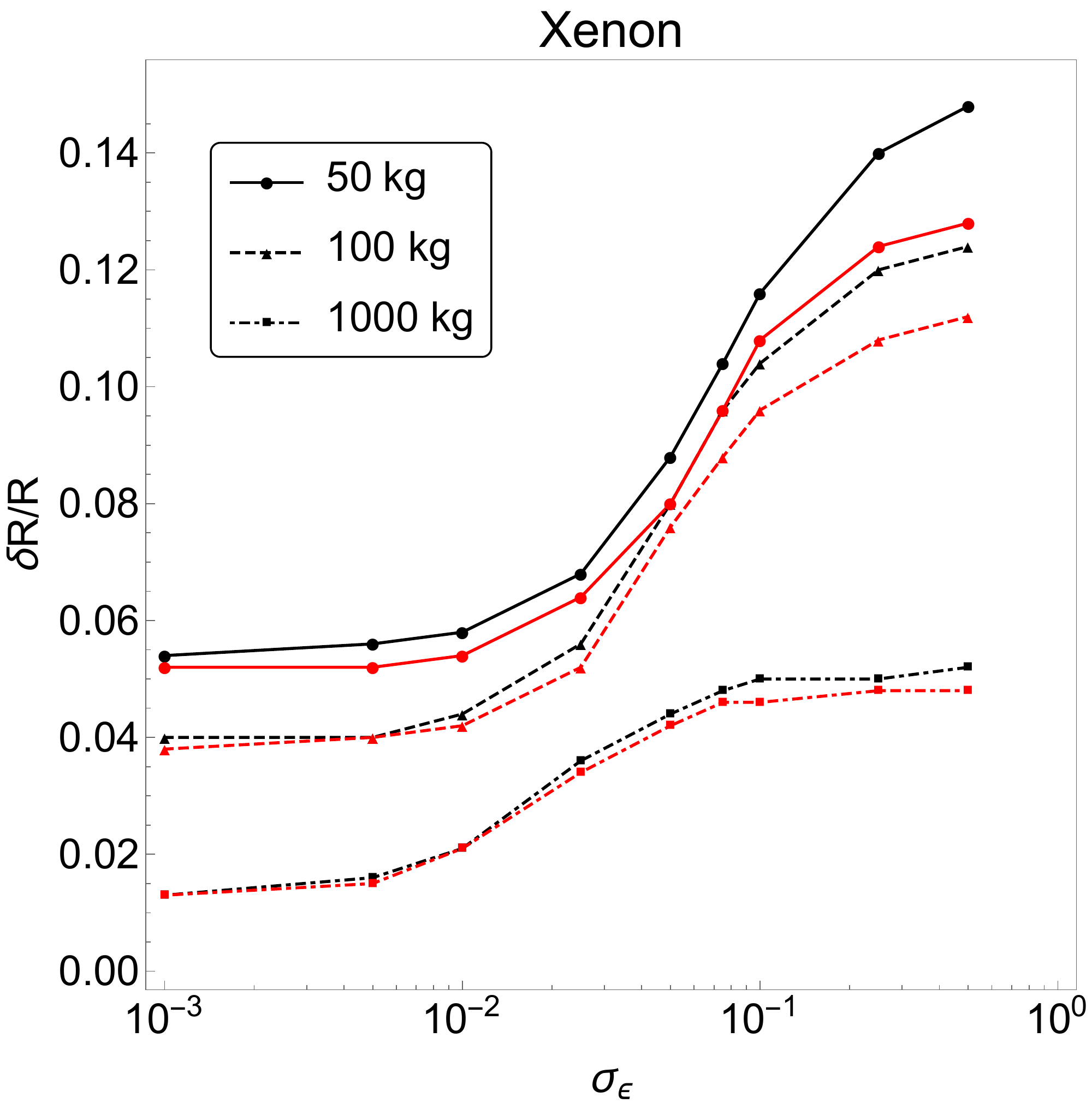}\vspace{-0.2cm}
\caption{\label{epsilon}Expected sensitivity as a function of the QF uncertainty}
\end{figure}
In Fig. \ref{sensitivity} it is reported the expected sensitivity on the neutron radius $R_n$ as a function of the detector mass, taking into account the uncertainty on the QF (assuming $\sigma_\epsilon=0.1$, solid curves) and assuming a perfect knowledge of the QF (dashed curves).
In Fig. \ref{epsilon}, instead, it is shown the expected sensitivity for a fixed detector mass, as a function of the uncertainty on the QF.
While the low-energy threshold of the detector can significantly change the total number of expected events, it does not  affect so much the sensitivity to the neutron radius. In Fig. \ref{spectrum} (left panel) it is shown the CE$\nu$NS expected spectrum for a liquid Argon detector, using the Helm model to compute the form factor $F(Q^2)$ (solid curve) and assuming $F(Q^2)=1$ (dashed curve): it is possible to see that most of the information on the form factor comes from the high-energy part of the spectrum; as a consequence, the low-energy threshold has little effect on the sensitivity to the neutron radius. In Fig. \ref{spectrum} (right panel) it is shown the expected sensitivity for a liquid Argon detector (1 ton) as a function of the low-energy threshold.
\section{Model-Independent Approach}
\label{MIapp}
It is possible to obtain a model-independent estimation of the neutron distribution by writing the form factor $F(Q^2)$ as a Taylor series of $Q^{2}$ (see also \cite {Patton}); each $Q^{2n}$ term will be multiplied by a factor proportional to the $2n$-th momenta of the neutron distribution, $\langle R^{2n}\rangle$. 
\[
F(Q^2)=1-\langle R^2\rangle Q^2/3!+\langle R^4\rangle Q^4/5!+\dots
\]
In this way it is possible to determine directly the momenta of the neutron distribution from the experimental data, without rely on any specific model. We calculated the 1-, 2- and 3-$\sigma$'s regions in the $\langle R^2\rangle-\langle R^4\rangle$ plane, taking into account also the $Q^6$ term and treating $\langle R^{6}\rangle$ as a pull parameter (the plots can be found in \cite{mio}). Also in this case, the leading source of uncertainty is the systematic error on the QF.
%%%%%%%%%%%%%%%%%%%%%%%%%%%%%%%%%%%%%%%%%%%%%%%%%%%%%%%%%%%%%%%%%%%%%%%%%
%%
%%   use this format to include an .eps figure into your paper
%%
%%%%%%%%%%%%%%%%%%%%%%%%%%%%%%%%%%%%%%%%%%%%%%%%%%%%%%%%%%%%%%%%%%%%%%%%%%%

\end{document}

%% file: econfmacros.tex
\def\beq{\begin{equation}}
\def\eeq#1{\label{#1}\end{equation}}
\def\eeqn{\end{equation}}

\def\beqa{\begin{eqnarray}}
\def\eeqa#1{\label{#1}\end{eqnarray}}
\def\eeqan{\end{eqnarray}}

\let\bar=\overbar

\def\Dslash{\not{\hbox{\kern-4pt $D$}}}
\def\dslash{\not{\hbox{\kern-2pt $\del$}}}

\def\msb{{\bar{\ssstyle M \kern -1pt S}}}